\documentclass{aastex} 

\usepackage{url}
\usepackage{lscape}
\usepackage{latexsym}
\usepackage{graphics}
\usepackage{graphicx}
\usepackage{enumitem}
\usepackage{amsmath}
\usepackage{mathptmx}
\usepackage{latexsym}
\usepackage[colorlinks,citecolor=blue,urlcolor=blue,linkcolor=blue]{hyperref}
\usepackage{fix-cm}
\usepackage{bigstrut}
\usepackage{booktabs}
\usepackage{array}

\shorttitle{Compact stars}
\shortauthors{Thomas and Pandya}

\begin{document}

\title{Compact stars on pseudo-spheroidal spacetime compatible with observational data}

\author{V. O. Thomas}
\affil{Department of Mathematics, Faculty of Science, The Maharaja Sayajirao University of Baroda, Vadodara 390 002, India}
\email{votmsu@gmail.com}

\author{D. M. Pandya} 
\affil{Department of Mathematics and Computer Science, Pandit Deendayal Petroleum University, Raisan, Gandhinagar 382 007, India}
\email{dishantpandya777@gmail.com}
 
\begin{abstract}

A new class of solutions for Einstein's field equations representing a static spherically symmetric anisotropic distribution of matter is obtained on the background of pseudo-spheroidal spacetime. We have prescribed the bounds of the model parameters $ k $ and $ p_0 $ on the basis of the elementary criteria for physical acceptability, viz., regularity, stability and energy conditions. By taking the values of model parameters from the prescribed bounds, we have shown that our model is compatible with the observational data of a wide variety of compact stars like 4U 1820-30, PSR J1903+327, 4U 1608-52, Vela X-1, PSR J1614-2230, SMC X-4 and Cen X-3.

\end{abstract}


\section{Introduction}
\label{sec:1}

\noindent The study of compact objects in agreement with observational data has received wide attention among researchers. A number of superdense star models, compatible with observational data, have appeared in literature in the recent past (~\cite{Murad13a}, ~\cite {Murad13b,Murad13c,Murad14a}, ~\cite{Maurya15} \& ~\cite {Sharma13}). If spacetime admitting compact star models possess a definite three-space geometry, then it is a mathematically interesting problem also. The spheroidal spacetimes studied by ~\cite {Vaidya82}, ~\cite {Tikekar90} and the paraboloidal spacetime studied by ~\cite {Finch89}, ~\cite {Jotania05}, ~\cite {Sharma13} are examples of spacetimes with definite 3-space geometry. The superdense star models developed by ~\cite {Tikekar98} has pseudo-spheroidal geometry. A number of researchers used this spacetime for developing physically viable models of compact stars under different assumptions on the physical content. \\
  
Theoretical investigations of ~\cite {Ruderman72} and ~\cite {Canuto74} suggest that matter may not isotropic in ultra high density regime. After the publication of the work of ~\cite {Bowers74}, there has been a large number of models devoted to the study of anisotropic distribution of matter. ~\cite {Maharaj89} developed an anisotropic model with uniform density and ~\cite {Gokhroo94} gave a more realistic anisotropic model with non-uniform density. ~\cite {Tikekar99,Tikekar05}, ~\cite {Thomas05} developed superdense anisotropic distributions on pseudo-spheroidal spacetimes. ~\cite {Thomas07} studied non-adiabatic gravitational collapse of anisotropic distribution of matter accompanied by radial heat flux. ~\cite {Dev02,Dev03,Dev04} have studied the impact of anisotropy on the stability of stellar configuration. Anisotropic distributions of matter incorporating linear equation of state have been studied by ~\cite {Sharma07}, \& ~\cite {Thirukkanesh08}. ~\cite {Komathiraj07} have studied charged distribution using linear equation of state. ~\cite {Sunzu14} studied charged anisotropic quark stars using linear equation of state. Anisotropic distributions of matter incorporating quadratic equation of state have been given by ~\cite {Feroze11} \& ~\cite {Maharaj12}. ~\cite {Varela10} used linear and non-linear equations of state for describing charged anisotropic distributions of matter. ~\cite {Paul11} have shown, in the MIT bag model of quark stars, that anisotropy can affect the bag constant. Polytropic equations of state has been used by ~\cite {Thirukkanesh12} \& ~\cite {Maharaj13b}. ~\cite {Malaver113,Malaver213,Malaver14} and ~\cite {Thirukkanesh14} have used modified Van der Waals equation of state for describing anisotropic charged compact stars.\\

Recently ~\cite{Pandya14} have developed anisotropic models of compact stars compatible with observational data by generalizing ~\cite {Finch89} ansatz. The anisotropic stellar model given by ~\cite {Sharma13} is a subclass of the model of ~\cite {Pandya14}. This model accommodates the observational data of a variety of compact objects recently studied by researchers. In the present article, we have obtained a new class of anisotropic stellar model of compact objects on the background of pseudo-spheroidal spacetimes. The physical parameter $ p_0 $ and geometric parameter $ k $ appearing in the model are restricted as a result of various physical acceptability conditions imposed on the model. Another geometric parameters $ R $ of the model plays the role of the radius of the spherical distribution of matter. It is found that our model yields values of different physical quantities that are in good agreement with the most recently available observational data of compact objects (~\cite{Gangopadhyay13}) like 4U 1820-30, PSR J1903+327, 4U 1608-52, Vela X-1, PSR J1614-2230, SMC X-4 and Cen X-3. \\

We have organized the paper as follows: In section ~\ref{sec:2}, we have solved the field equations and obtained restrictions on the model parameters using various physical requirements and energy conditions. The bounds for the model parameters $ k $ and $ p_0 $ are obtained in section ~\ref{sec:3}. In section ~\ref{sec:4}, we have shown that our model is compatible with recent observational data of a number of compact objects (~\cite{Gangopadhyay13}). The main results obtained in the present work is discussed in section ~\ref{sec:5}.
\section{Spacetime metric}
\label{sec:2}

A three-pseudo spheroid immersed in four-dimensional Euclidean space has the Cartesian equation

\begin{eqnarray}
 \frac{u^2}{b^2} - \frac{x^2 + y^2 + z^2}{R^2} = 1. \nonumber
 \label{three-pseudo spheroid equation}
\end{eqnarray}

The sections $ u = constant $ are spheres of real or imaginary radius according as $ u^2 > b^2 $ or $ u^2 < b^2, $ while the sections $ x = const,~y = const $, and $ z = const $ are respectively, hyperboloids of two sheets. \\ 
On taking the parametrization 

\begin{eqnarray}\label{parametrization}
 \nonumber x & = & R sinh\lambda sin\theta cos\phi \\ \nonumber
 y & = & R sinh\lambda sin\theta sin\phi \\ \nonumber
 z & = & R sinh\lambda cos\theta \\
 u & = & b cosh\lambda
\end{eqnarray}
the Euclidean metric
\begin{equation}
 d\sigma^2 = dx^2 + dy^2 + dz^2 + du^2 \nonumber
 \label{sigma}
\end{equation}
takes the form
\begin{equation}
 d\sigma^2 = \frac{1 + k \frac{r^2}{R^2}}{1 + \frac{r^2}{R^2}} dr^2 + r^2 d\theta^2 + r^2 sin^2 \theta d\phi^2
 \label{sigma in spherical coordinates}
\end{equation}

where $ k = 1 + \frac{b^2}{R^2} $ and $ r = R sinh\lambda $. The metric (\ref{sigma in spherical coordinates}) is regular for all points with $ k > 1 $ and call pseudo-spheroidal metric (\cite{Tikekar98}). \\
We take the interior metric describing the anisotropic matter distribution in the form
\begin{equation}
ds^2 = e^{\nu (r)} dt^2 - \frac{1 + k \frac{r^2}{R^2}}{1 + \frac{r^2}{R^2}}dr^2 - r^2 d\theta^2 - r^2 sin^2 \theta d\phi^2 ,
\label{a}
\end{equation}
\noindent where, $ k $, $ R $ are geometric parameters and $ k > 1 $. This spacetime, generally known as pseudo-spheroidal spacetime, has been studied by many researchers (\cite {Tikekar98,Tikekar99,Tikekar05,Thomas05,Thomas07,Paul11,Chattopadhyay10,Chattopadhyay12}). \\
\noindent Following ~\cite {Maharaj89}, we write the energy-momentum tensor for anisotropic matter distribution in the form
\begin{equation}
T_{ij} = \left(\rho + p \right) u_i u_j - p g_{ij} + \pi_{ij},
\label{b}
\end{equation}
\noindent where, $ \rho, p $ and $ u_i $ denote the proper density, fluid pressure and unit four-velocity of the fluid, respectively. \\
\noindent The anisotropic stress-tensor $\pi_{ij}$ is given by
\begin{equation}
\pi_{ij} = \sqrt{3}S \left[C_iC_j - \frac{1}{3}(u_iu_j - g_{ij})\right],
\label{c}
\end{equation}
\noindent where, $ C^i = (0,-e^{-\lambda/2},0,0) $ is a radial vector and $ S = S(r) $ denotes the magnitude of the anisotropic stress. \\
\noindent The non-vanishing components of the energy-momentum tensor are given by
\begin{equation}
T_0^0 = \rho, ~~~ T_1^1 = - \left(p + \frac{2S}{\sqrt{3}}\right), ~~~ T_2^2 = T_3^3  =  -\left(p - \frac{S}{\sqrt{3}}\right).
\label{d}
\end{equation}
\noindent Hence the radial and transverse pressures are given by
\begin{eqnarray}
p_r &=& -T_1^1 = \left(p + \frac{2S}{\sqrt{3}}\right),\label{e}\\
p_\perp &=& -T_2^2 = \left(p - \frac{S}{\sqrt{3}}\right).\label{f}
\end{eqnarray}
Then the magnitude of the anisotropic stress has the form 
\begin{equation}
S = \frac{p_r - p_\perp}{\sqrt{3}}.
\label{g}
\end{equation}
\noindent The physical and geometric variables, related through Einstein's field equations 
\begin{equation}
R_{ij} - \frac{1}{2}Rg_{ij} = 8\pi T_{ij},
\label{h} 
\end{equation}
\noindent are to be determined from the following set of three equations:
\begin{eqnarray}
8\pi \rho &=& \frac{1 - e^{-\lambda}}{r^2} + \frac{e^{-\lambda} {\lambda}^\prime}{r},\label{i}\\
8\pi p_r &=& \frac{e^{-\lambda} - 1}{r^2} - \frac{e^{-\lambda}{\nu}^\prime}{r},\label{j}\\
8\pi p_\perp &=& e^{-\lambda} \left[\frac{\nu^{\prime\prime}}{2} + \frac{{\nu^\prime}^2}{4} - \frac{\nu^\prime \lambda^\prime}{4} + \frac{\nu^\prime - \lambda^\prime}{2r}\right],\label{k}
\end{eqnarray}
\noindent where a prime denotes a differentiation with respect to $ r $. The equations (\ref{i}) -- (\ref{k}) can be couched in the form
\begin{eqnarray}
e^{-\lambda} &=& 1 - \frac{2m}{r},\label{l}\\
\left(1 - \frac{2m}{r}\right)\nu^{\prime} &=& 8\pi p_r r + \frac{2m}{r^2},\label{m}\\
-\frac{4}{r} (8\pi \sqrt{3}S) &=& (8\pi\rho + 8\pi p_r)\nu^{\prime} + 2 (8\pi p_r^{\prime}),\label{n}
\end{eqnarray}
\noindent where
\begin{equation}
m(r) = 4\pi \int\limits_{0}^{r} u^2 \rho(u) du.
\label{o}
\end{equation}
\noindent The energy-density $ \rho $ and the mass $ m $ within the radius $ r $ have expressions
\begin{equation}
8 \pi \rho = \frac{k-1}{R^2} \frac{\left(3 + k \frac{r^2}{R^2}\right)}{\left(1 + k \frac{r^2}{R^2}\right)^2},
\label{p}
\end{equation}
\begin{equation}
m(r) = \frac{R}{2} \frac{(k-1)\frac{r^2}{R^2}}{1 + \frac{k r^2}{R^2}}.
\label{q}
\end{equation}
\noindent It can be easily obtained from equation (\ref{p}) that
\begin{equation}
8\pi \rho' = - \frac{2k(k-1)r}{R^4} \frac{\left(5 + k \frac{r^2}{R^2}\right)}{\left(1 + k \frac{r^2}{R^2}\right)^3} < 0,
\label{r} 
\end{equation} 
\noindent indicating that the density decreases radially outward.\\
\noindent In order to obtain the metric potential $ \nu $, we assume an expression for $ p_r $ in equation (\ref{m}), in the form
\begin{equation}
8\pi p_r = \frac{p_0}{R^2} \frac{\left(1 - \frac{r^2}{R^2}\right)\left(1 + \frac{r^2}{R^2}\right)}{\left(1 + k \frac{r^2}{R^2}\right)^2}.
\label{s}
\end{equation}
\noindent The radial pressure in the present form vanishes at $ r = R $ and takes the value $ \frac{p_0}{R^2} $ at the centre $ r = 0. $ It is non-negative for all values of $ r $ in the range $ 0 \leq r \leq R. $ Further, on differentiating equation (\ref{s}) with respect to $ r $, we get 
\begin{equation}
8 \pi p_r' = - \frac{4p_0r\left(k + \frac{r^2}{R^2}\right)}{R^4 \left(1 + k \frac{r^2}{R^2}\right)^3} < 0,
\label{t} 
\end{equation}
\noindent indicating that the pressure $ p_r $ decreases radially outward. Since $ p_r(r=R)=0, $ the geometric parameter $ R $ takes the role of the boundary radius of the distribution. With this choice of $ p_r $, equation (\ref{m}) can be integrated to obtain $ \nu $ in the form
\begin{eqnarray}
e^\nu = A \left(1 + k \frac{r^2}{R^2}\right)^\frac{p_0 (k+1)}{2k^2}\left(1 + \frac{r^2}{R^2}\right)^\frac{k-1}{2} \nonumber \\
exp \left\lbrace\frac{-p_0}{2k^2}\left(1 + k \frac{r^2}{R^2}\right)\right\rbrace
\label{u}
\end{eqnarray}
\noindent where $ A $ is a constant of integration. \\
\noindent Therefore, the spacetime metric takes the explicit form
\begin{eqnarray}
ds^2 = A \left(1 + k \frac{r^2}{R^2}\right)^\frac{p_0 (k+1)}{2k^2}\left(1 + \frac{r^2}{R^2}\right)^\frac{k-1}{2} \nonumber \\
\times exp \left\lbrace\frac{-p_0}{2k^2}\left(1 + k \frac{r^2}{R^2}\right)\right\rbrace dt^2 - \frac{1 + k \frac{r^2}{R^2}}{1 + \frac{r^2}{R^2}}dr^2 - r^2 d\theta^2 - r^2 sin^2 \theta d\phi^2. 
\label{v}
\end{eqnarray} 
\noindent The constant of integration $ A $ can be obtained by matching the interior spacetime metric (\ref{a}) with the Schwarzschild exterior metric
\begin{equation}
 ds^2 = \left(1 - \frac{2m}{r}\right) dt^2 - \left(1 - \frac{2m}{r}\right)^{-1}dr^2 - r^2 d\theta^2 - r^2 sin^2 \theta d\phi^2
 \label{mm}
\end{equation}
\noindent across the boundary $ r = R. $ This gives 
\begin{equation}
R = \frac{2M(k+1)}{(k-1)},
\label{w}
\end{equation} 
\noindent and
\begin{equation}
A = \frac{2}{k+1}\left(\frac{e}{k+1}\right)^{\frac{(k+1)p_0}{2 k^2}}2^{-\left(\frac{k-1}{2}\right)}. \\
\label{x}  
\end{equation}
\noindent The expression for anisotropy is now readily available by substituting for $ p_r $, $ p_r' $ and $ \nu' $ in the equation (\ref{n}).
\begin{eqnarray}
8\pi\sqrt{3}S=\frac{r^2}{R^2} \left(\frac{2 p_0\left(k+\frac{r^2}{R^2}\right)}{R^2 \left(1 + \frac{k r^2}{R^2}\right)^3} - \frac{B(r)C(r)}{R^2\left(1 + k \frac{r^2}{R^2}\right)}\right),
\label{y}
\end{eqnarray}
\noindent where 
\begin{equation}
B(r) = \left[\frac{p_0 \left(1-\frac{r^2}{R^2}\right)}{4 \left(1 + \frac{k r^2}{R^2}\right)^4}+\frac{k-1}{4 \left(1 + \frac{r^2}{R^2}\right)}\right],
\end{equation}
\begin{equation}
C(r) = \left[(k-1)\left(3 + k\frac{r^2}{R^2}\right)+ p_0 \left(1-\frac{r^4}{R^4}\right)\right].
\end{equation}
\noindent It is easy to see that $ S $ vanishes at origin $ r = 0, $ which is a desired requirement for anisotropic distributions (~\cite {Murad13a}, ~\cite {Murad13b,Murad13c,Murad14a} \& ~\cite {Bowers74}).\\
\noindent The expression for transverse pressure 
\begin{equation}
8 \pi p_\perp = 8 \pi p_r - 8 \pi \sqrt{3} S
\label{z} 
\end{equation}
\noindent can be obtained using equations (\ref{s}) and (\ref{y}).\\
\noindent Moreover, the condition $ p_\perp \geq 0 $ will lead to the following inequality at $ r = R $
\begin{equation}
p_0 \leq \frac{1}{16} (k-1)^2 (k+3),
\label{ii}
\end{equation}
\noindent whereas at $ r = 0 $, the condition is evidently satisfied. \\
\noindent The expressions for $ \frac{dp_r}{d\rho} $ and $ \frac{dp_\perp}{d\rho} $ are given by
\begin{equation}
\frac{dp_r}{d\rho} = \frac{2p_0}{k(k-1)}\frac{\left(k + \frac{r^2}{R^2}\right)}{\left(5 + k \frac{r^2}{R^2}\right)},
\label{aa} 
\end{equation}
\begin{equation}
\frac{dp_\perp}{d\rho} = \frac{dp_r}{d\rho} - \sqrt{3}\frac{dS}{d\rho}.
\label{bb}
\end{equation}
\noindent The conditions $ 0 \leq \frac{dp_r}{d\rho} \leq 1 $ and $ 0 \leq \frac{dp_\perp}{d\rho} \leq 1 $ at $ r = 0 $, respectively, give the inequalities
\begin{equation}
0 \leq p_0 \leq \frac{5}{2} (k-1),
\label{cc}
\end{equation}
\noindent and
\begin{equation}
0 \leq p_0 \leq \frac{k (k-1)(k+5)}{2(k+1)}.
\label{dd}
\end{equation}
\noindent Similarly the above conditions at $ r = R $, respectively, give
\begin{equation}
2(3k+1)-\sqrt{33k^2+30k+1} \leq p_0 \leq 2(3k+1)-\sqrt{13k^2+50k+1},
\label{ee}
\end{equation}
\noindent and
\begin{equation}
0< p_0 \leq \frac{-k^4+12k^3+78k^2-92k+3}{8k^2-24k+80}.
\label{ff}
\end{equation}
\noindent The adiabatic index 
\begin{equation}
\Gamma = \frac{\rho + p_r}{p_r} \frac{dp_r}{d\rho}
\label{gg}
\end{equation}
\noindent has the explicit expression 
\begin{equation}
\Gamma = \frac{2 \left(k + \frac{r^2}{R^2}\right)C(r)}{k(k-1)\left(5 + k \frac{r^2}{R^2}\right)\left(1 - \frac{r^4}{R^4}\right)}.
\label{hh}
\end{equation}
\noindent The necessary condition for the model to represent a relativistic star is that $ \Gamma > \frac{4}{3} $ throughout the star. $ \Gamma > \frac{4}{3} $ at $ r = 0 $ impose a condition on $ p_0 $, viz., 
\begin{equation}
p_0 > \frac{k-1}{3}.
\label{jj}
\end{equation}
\noindent The strong energy condition $ \rho - p_r - 2 p_\perp \geq 0 $ at $ r = 0 $ and $ r = R, $ respectively, give the following two inequalities 
\begin{equation}
p_0 \leq k-1
\label{kk}
\end{equation}
\noindent and
\begin{equation}
p_0 \geq \frac{(k+3)(k-1)(k-5)}{16}
\label{ll}
\end{equation}
\noindent In order to obtain a valid range for the parameters $ p_0 $ and $ k $, we have to consider the inequalities~(\ref{cc}) --~(\ref{ff}) and~(\ref{hh}) --~(\ref{ll}) simultaneously. 

\section{Bounds for Model Parameters}
\label{sec:3}

\noindent The pseudo-spheroidal space-time model developed for anisotropic matter distribution contains a physical parameter $ p_0 $ related to the central pressure and two geometric parameters, viz., $ R $ and $ k $. Since $ p_r (r = R) = 0 $, the free parameter $ R $ represents the radius of the distribution. The bounds for the other two parameters $ p_0 $ and $ k $ are to be determined by the following requirements a physically acceptable model is expected to satisfy in its region of validity, $ 0 \leq r \leq R $. \\

\begin{enumerate}[label=\textbf{\arabic*})]
 
 \item $ \rho (r) \geq 0, ~~ p_r (r) \geq 0, ~~ p_\perp (r) \geq 0 $; \\
 
 \item $ \rho(r) - p_r (r) - 2 p_\perp (r) \geq 0 $ ; \\
 
 \item $ \frac{d\rho (r)}{dr} < 0, ~~ \frac{dp_r (r)}{dr} < 0 $; \\
 
 \item $ 0 \leq \frac{dp_r}{d\rho} \leq 1, ~~ 0 \leq \frac{dp_\perp}{d\rho} \leq 1 $ ; \\
 
 \item The adiabatic index $ \Gamma(r) > \frac{4}{3}. $ \\ 
 
\end{enumerate}
\noindent The conditions $ \rho (r) \geq 0, p_r (r) \geq 0, \frac{d\rho(r)}{dr} < 0, \frac{dp_r(r)}{dr} < 0 $ are automatically satisfied by equations (\ref{p}), (\ref{s}), (\ref{r}), (\ref{t}). \\
\noindent We have displayed in Table ~\ref{tab:1} the bounds on $ p_0 $ in terms of the parameter $ k $ at the centre and on the boundary.

\begin{table}[hbtp]
\caption{Bounds for $ p_0 $.}
\label{tab:1}      
\renewcommand{\arraystretch}{2}
\begin{tabular}{cccc}

\textbf{Physical requirements} & \textbf{at $ r = 0 $} & \textbf{at $ r = R $} \\ \hline

$ \rho - p_r - 2p_\perp \geq 0  $ & $ p_0 \leq k-1 $ & $ p_0 \geq \frac{(k+3)(k-1)(k-5)}{16} $ \\ 

$ 0 \leq \frac{dp_r}{d\rho} \leq 1 $ & $ 0 \leq p_0 \leq \frac{5}{2}(k-1) $ & $ {2(3k+1)-\sqrt{(33k^2 + 30k + 1)}} \leq p_0 $ \\
 & & $ \leq {2(3k+1)-\sqrt{(13k^2 + 50k + 1)}} $ \\

$ 0 \leq \frac{dp_\perp}{d\rho} \leq 1 $ & $ 0 \leq p_0 \leq \frac{k(k-1)(k+5)}{2(k+1)} $ & $ 0 \leq p_0 \leq \frac{-k^4 + 12 k^3 + 78 k^2 - 92 k + 3}{8k^2 - 24 k +80}$ \\ 

$ \Gamma(r) \geq \frac{4}{3} $ & $ p_0 > \frac{k-1}{3} $ & Automatically satisfied \\ \hline

\end{tabular} 
\end{table}

\begin{figure}[h]
\resizebox{0.65\textwidth}{!}{%
  \includegraphics{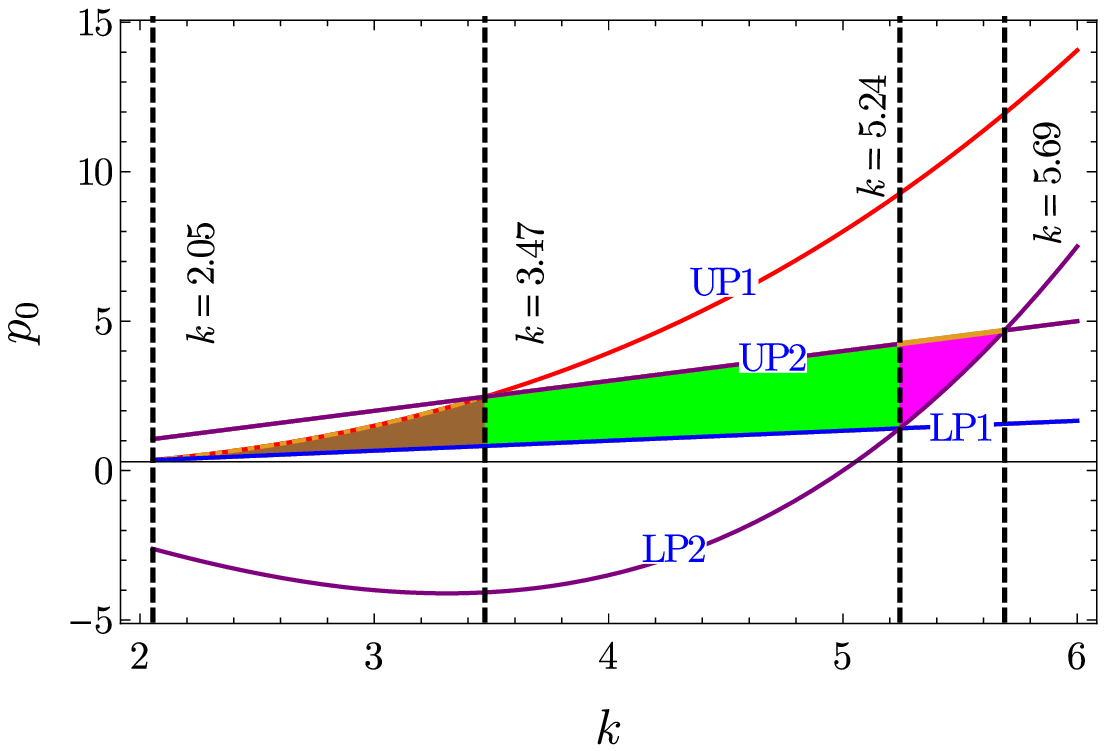}																	
}
\caption{Permissible values of $ p_0 $ and $ k $}
\label{fig:1}       
\end{figure}

\noindent We have displayed the numerical values of the lower and upper bounds of $ p_0 $ for different values of $ k > 1 $ in Table ~\ref{tab:2}. We have considered the maximum of all lower limits of $ p_0 $ and minimum of all its upper limits. The admissible values of $ k $ are those for which minimum of upper limit minus maximum of lower limit is positive. This condition restricts the values of $ k $ in the range $ (2.05, 5.69). $ It is further observed that for $ 2.05 < k \leq 3.47,~ 3.47 \leq k \leq 5.24, $ and $ 5.24 \leq k < 5.69,~p_0 $ satisfies, respectively, the inequalities $ \frac{k-1}{3} < p_0 \leq \frac{(k+3)(k-1)^2}{16},~\frac{k-1}{3} \leq p_0 \leq k-1 $ and $ \frac{1}{6}(k+3)(k-1)(k-5) \leq p_0 < k-1 $. The shaded region in Figure ~\ref{fig:1} gives the permissible values of $ k $ and $ p_0. $ Any values of $ k $ and $ p_0 $ outside this region may violate one or other of the physical requirements of the model.

\begin{table}[hbtp]
\caption{Permissible values of $ k $ and $ p_0. \newline ~~ \text{Here,} ~~ LP1 = \frac{k-1}{3}, ~~ LP2 = \frac{(k+3)(k-1)(k-5)}{6}, ~~ LP3 = {2(3k+1)-\sqrt{(33k^2 + 30k + 1)}}, \newline ~~ UP1 = \frac{(K+3)(k-1)^2}{16}, ~~ UP2 = k-1, ~~ UP3 = \frac{5(k-1)}{2}, \newline ~~ UP4 = \frac{k(k-1)(k+5)}{2(k+1)}, ~~ UP5 = {2(3k+1)-\sqrt{(13k^2 + 50k + 1)}}, ~~ UP6 = \frac{-k^4 + 12 k^3 + 78 k^2 - 92 k + 3}{8k^2 - 24 k +80}. $}
\label{tab:2}
\begin{tabular}{cccccccccccccc}\hline
\multicolumn{1}{c}{$ k $} & \multicolumn{3}{c}{\textbf{Lower Limit for $ p_0 $}} & \multicolumn{1}{c}{$ \textbf{Max} $} & \multicolumn{6}{c}{\textbf{Upper Limit for $ p_0 $}} & \multicolumn{1}{c}{$ \textbf{Min} $} & \multicolumn{1}{c}{$ \textbf{Min - Max} $} \\ \cline{2-4}\cline{6-11}
\multicolumn{1}{c}{} & \multicolumn{1}{c}{$ LP1 $} & \multicolumn{1}{c}{$ LP2 $} & \multicolumn{1}{c}{$ LP3 $} & \multicolumn{1}{c}{} & \multicolumn{1}{c}{$ UP1 $} & \multicolumn{1}{c}{$ UP2 $} & \multicolumn{1}{c}{$ UP3 $} & \multicolumn{1}{c}{$ UP4 $} & \multicolumn{1}{c}{$ UP5 $} & \multicolumn{1}{c}{$ UP6 $} & \multicolumn{1}{c}{} & \multicolumn{1}{c}{} \bigstrut \\ \hline

2 & \textbf{0.33} & -2.50 & 0.11 & \textbf{0.33} & \textbf{0.31} & 1 & 2.5 & 2.33 & 1.63 & 3.30 & \textbf{0.31} & -0.02 \\ 
\textbf{2.05} & \textbf{0.35} & -2.61 & 0.12 & \textbf{0.35} & \textbf{0.35} & 1.05 & 2.625 & 2.49 & 1.72 & 3.54 & \textbf{0.35} & \textbf{0.00} \\
2.1 & \textbf{0.37} & -2.71 & 0.12 & \textbf{0.37} & \textbf{0.39} & 1.1 & 2.75 & 2.65 & 1.82 & 3.78 & \textbf{0.39} & 0.02 \\
2.4 & \textbf{0.47} & -3.28 & 0.18 & \textbf{0.47} & \textbf{0.66} & 1.4 & 3.5 & 3.66 & 2.40 & 5.32 & \textbf{0.66} & 0.19 \\ 
2.8 & \textbf{0.60} & -3.83 & 0.26 & \textbf{0.60} & \textbf{1.17} & 1.8 & 4.5 & 5.17 & 3.21 & 7.40 & \textbf{1.17} & 0.57 \\ 
3 & \textbf{0.67} & -4.00 & 0.30 & \textbf{0.67} & \textbf{1.50} & 2 & 5 & 6.00 & 3.63 & 8.40 & \textbf{1.50} & 0.83 \\
\textbf{3.1} & \textbf{0.70} & -4.06 & 0.32 & \textbf{0.70} & \textbf{1.68} & 2.1 & 5.25 & 6.43 & 3.84 & 8.88 & \textbf{1.68} & \textbf{0.98} \\
3.2 & \textbf{0.73} & -4.09 & 0.35 & \textbf{0.73} & \textbf{1.88} & 2.2 & 5.5 & 6.87 & 4.05 & 9.35 & \textbf{1.88} & 1.14 \\
3.4 & \textbf{0.80} & -4.10 & 0.39 & \textbf{0.80} & \textbf{2.30} & 2.4 & 6 & 7.79 & 4.48 & 10.23 & \textbf{2.30} & 1.50 \\ 
\textbf{3.47} & \textbf{0.82} & -4.08 & 0.40 & \textbf{0.82} & \textbf{2.47} & 2.47 & 6.175 & 8.12 & 4.63 & 10.53 & \textbf{2.47} & \textbf{1.64} \\
3.8 & \textbf{0.93} & -3.81 & 0.48 & \textbf{0.93} & 3.33 & \textbf{2.8} & 7 & 9.75 & 5.34 & 11.79 & \textbf{2.80} & 1.87 \\
4 & \textbf{1.00} & -3.50 & 0.52 & \textbf{1.00} & 3.94 & \textbf{3} & 7.5 & 10.80 & 5.78 & 12.46 & \textbf{3.00} & 2.00 \\
4.2 & \textbf{1.07} & -3.07 & 0.57 & \textbf{1.07} & 4.61 & \textbf{3.2} & 8 & 11.89 & 6.22 & 13.05 & \textbf{3.20} & 2.13 \\ 
4.4 & \textbf{1.13} & -2.52 & 0.62 & \textbf{1.13} & 5.35 & \textbf{3.4} & 8.5 & 13.02 & 6.66 & 13.58 & \textbf{3.40} & 2.27 \\
4.8 & \textbf{1.27} & -0.99 & 0.71 & \textbf{1.27} & 7.04 & \textbf{3.8} & 9.5 & 15.41 & 7.55 & 14.45 & \textbf{3.80} & 2.53 \\ 
5 & \textbf{1.33} & 0.00 & 0.76 & \textbf{1.33} & 8.00 & \textbf{4} & 10 & 16.67 & 8.00 & 14.80 & \textbf{4.00} & 2.67 \\
5.2 & \textbf{1.40} & 1.15 & 0.81 & \textbf{1.40} & 9.04 & \textbf{4.2} & 10.5 & 17.97 & 8.45 & 15.10 & \textbf{4.20} & 2.80 \\ 
\textbf{5.24} & \textbf{1.41} & 1.40 & 0.82 & \textbf{1.41} & 9.26 & \textbf{4.24} & 10.6 & 18.23 & 8.54 & 15.15 & \textbf{4.24} & \textbf{2.83} \\
5.4 & 1.47 & \textbf{2.46} & 0.85 & \textbf{2.46} & 10.16 & \textbf{4.4} & 11 & 19.31 & 8.90 & 15.35 & \textbf{4.40} & 1.94 \\
5.67 & 1.56 & \textbf{4.52} & 0.92 & \textbf{4.52} & 11.82 & \textbf{4.67} & 11.675 & 21.18 & 9.52 & 15.63 & \textbf{4.67} & 0.15 \\
\textbf{5.69} & 1.56 & \textbf{4.69} & 0.92 & \textbf{4.69} & 11.95 & \textbf{4.69} & 11.725 & 21.32 & 9.56 & 15.64 & \textbf{4.69} & \textbf{0.00} \\ 
5.71 & 1.57 & \textbf{4.85} & 0.93 & \textbf{4.85} & 12.08 & \textbf{4.71} & 11.775 & 21.46 & 9.61 & 15.66 & \textbf{4.71} & -0.14 \\ \hline
\end{tabular}
\end{table}

\pagebreak

\section{Compact Star Models}
\label{sec:4}

\noindent In order to validate the model, we examine our model with observational data. We have considered the pulsar 4U 1820-30 whose estimated mass and radius are $ 1.58 M_\odot $ and 9.1 km (~\cite {Gangopadhyay13}). If we set these values for mass and radius then from equation (\ref{w}) we obtain the value of $ k = 3.1 $ which is well inside the valid range for $ k $. Similarly assuming masses of some well studied compact stars like PSR J1903+327, 4U 1608-52, Vela X-1, PSR J1614-2230, SMC X-4 and Cen X-3, we have obtained the same radius calculated by ~\cite {Gangopadhyay13} for values of $ k $ in the valid range. The values of mass, radius, $ k $ and other relevant quantities like central density $ \rho_c, $ density at the boundary $ \rho_R $, the compactification parameter $ u $ and $ \frac{dp_r}{d\rho} $ at the centre for $ p_0 = 1.08, $ are shown in Table ~\ref{tab:3}. \\   

\begin{table}[h]
\caption{Estimation of physical values based on observational data for $ p_0 = 1.08 $.}
\label{tab:3}       
\begin{tabular}{llllllll}
\hline\noalign{\smallskip}
\textbf{STAR} & $ k $ & \textbf{$ M $} & \textbf{$ R $} & \textbf{$ \rho_c $} & \textbf{$ \rho_R $} & \textbf{$ u (=\frac{M}{R}) $} & $\left(\frac{dp_r}{d \rho}\right)_{r=0} $ \\
& & $ (M_\odot) $ & (Km) & (MeV fm{$^{-3}$}) & (MeV fm{$^{-3}$}) & \\
\noalign{\smallskip}\hline\noalign{\smallskip}
\textbf{4U 1820-30} 	  & 3.100 & 1.58  & 9.1   & 2290.97 & 277.12 & 0.256 & 0.206 \\
\textbf{PSR J1903+327} 	  & 3.176 & 1.667 & 9.438 & 2129.82 & 257.62 & 0.261 & 0.199 \\
\textbf{4U 1608-52} 	  & 3.458 & 1.74  & 9.31  & 2188.78 & 267.75 & 0.276 & 0.176 \\
\textbf{Vela X-1} 	  & 3.407 & 1.77  & 9.56  & 2075.80 & 251.08 & 0.273 & 0.179 \\
\textbf{PSR J1614-2230}   & 3.997 & 1.97  & 9.69  & 2020.48 & 244.39 & 0.300 & 0.144 \\
\textbf{SMC X-4}          & 2.514 & 1.29  & 8.831 & 2432.67 & 294.25 & 0.215 & 0.285 \\
\textbf{Cen X-3}          & 2.838 & 1.49  & 9.178 & 2252.20 & 272.42 & 0.239 & 0.235 \\
\noalign{\smallskip}\hline
\end{tabular} 

\end{table}

\begin{figure}[h]
\resizebox{0.62\textwidth}{!}{%
  \includegraphics{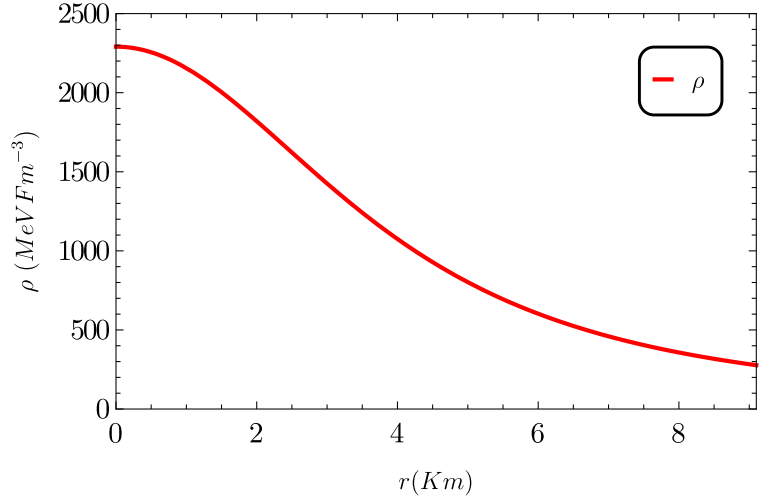}																	
}
\caption{Density Vs Radius}
\label{fig:2}       
\end{figure}

\begin{figure}[h]
\resizebox{0.62\textwidth}{!}{%
  \includegraphics{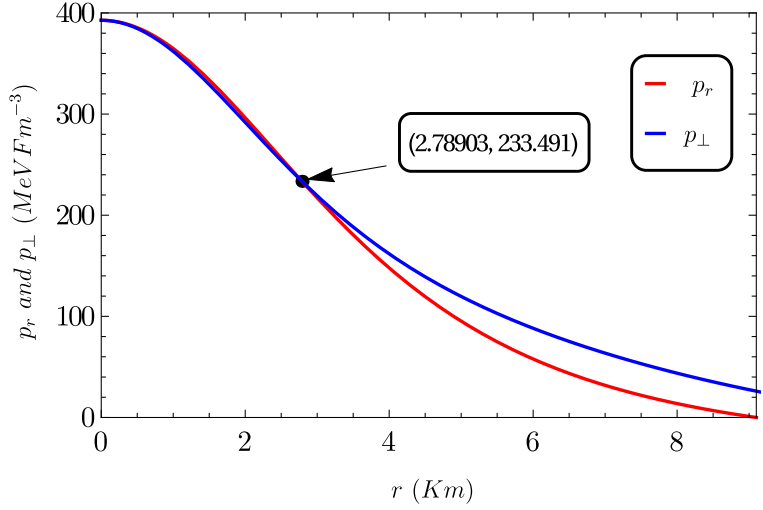}																	
}
\caption{Radial and Transverse Pressures Vs Radius}
\label{fig:3}       
\end{figure}

\begin{figure}[h]
\resizebox{0.62\textwidth}{!}{%
  \includegraphics{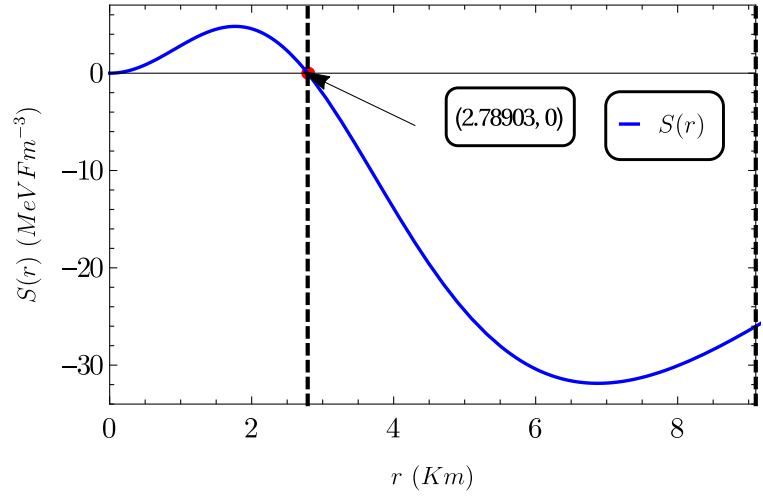}																		
}
\caption{Anisotropy Vs Radius}
\label{fig:4}       
\end{figure}

\begin{figure}[h]
\resizebox{0.62\textwidth}{!}{%
  \includegraphics{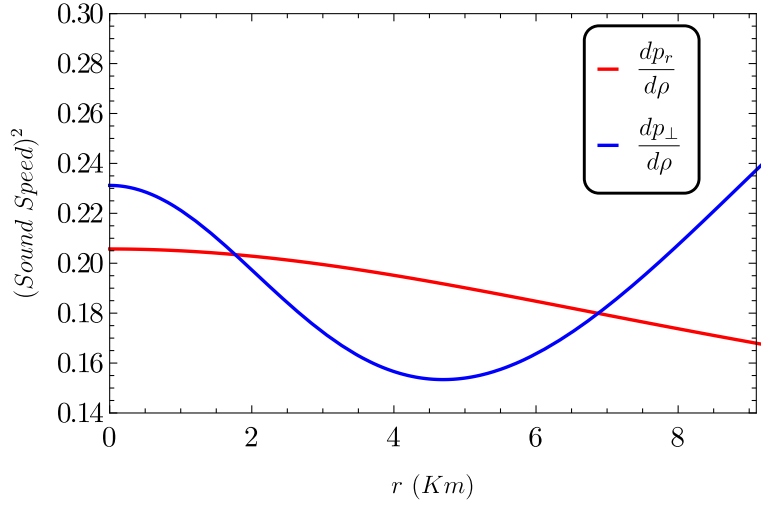}																	
}
\caption{$ (Sound~Speed)^2 $ Vs Radius}
\label{fig:5}       
\end{figure}

\begin{figure}[h]
\resizebox{0.62\textwidth}{!}{%
  \includegraphics{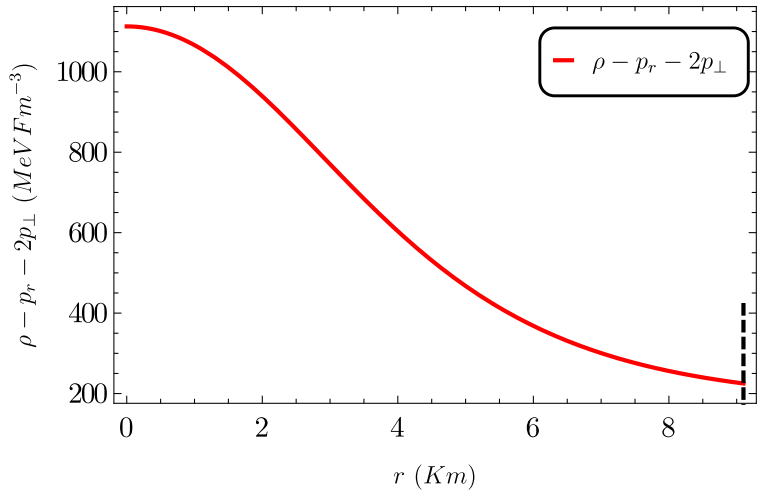}																	
}
\caption{Strong Energy Condition Vs Radius}
\label{fig:6}       
\end{figure}

\begin{figure}
\resizebox{0.62\textwidth}{!}{%
  \includegraphics{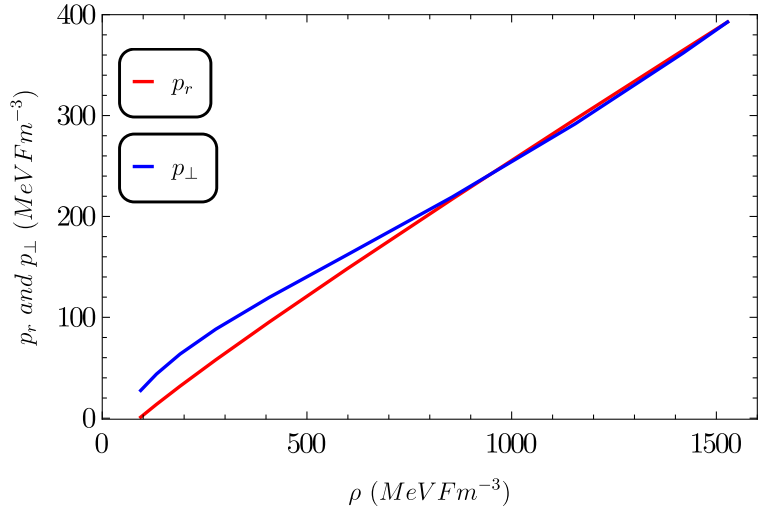}																	
}
\caption{Equation of State}
\label{fig:7}       
\end{figure}

\begin{figure}
\resizebox{0.62\textwidth}{!}{%
  \includegraphics{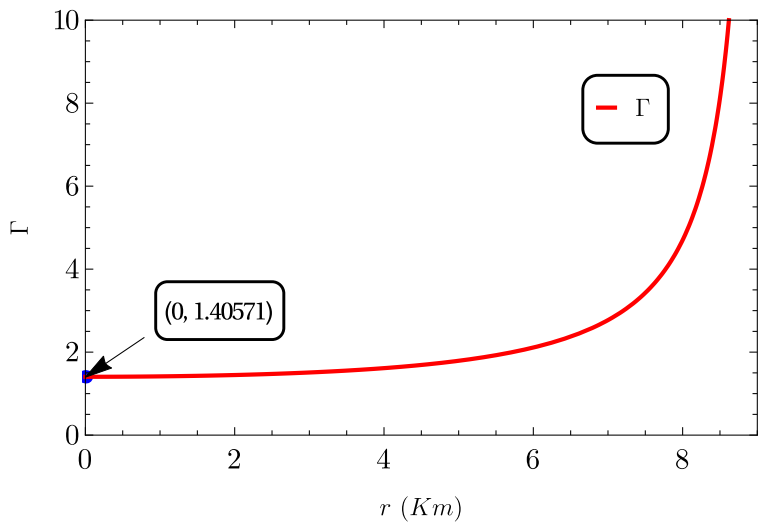}																	
}
\caption{Adiabatic Index Vs Radius}
\label{fig:8}      
\end{figure}

\begin{figure}
\resizebox{0.62\textwidth}{!}{%
  \includegraphics{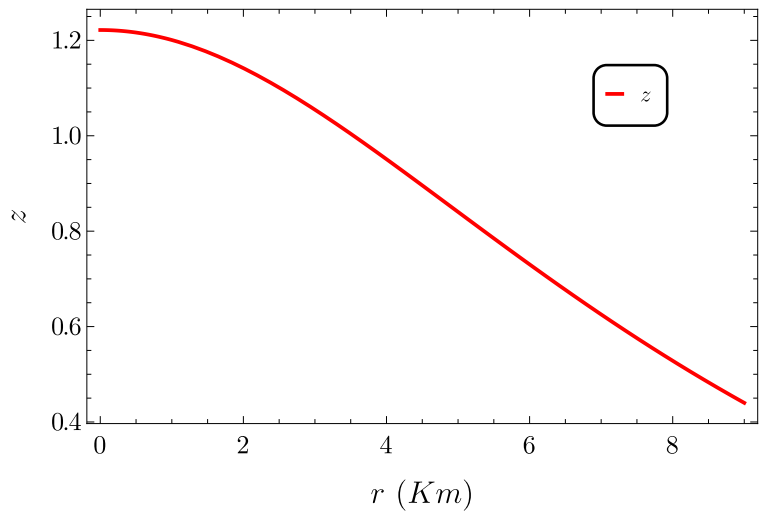}																	
}
\caption{Gravitational Red Shift}
\label{fig:9}      
\end{figure}

\noindent In order to examine the nature of various physical quantities throughout the distribution, we have considered a particular star 4U 1820-30 for which mass $ M = 1.58 M_\odot, $ radius $ R = 9.1 km, $ the physical parameter $ p_0 = 1.08 $ and the geometric parameter $ k = 3.1 $. We have shown the variation of density and pressures in Figure ~\ref{fig:2} and Figure ~\ref{fig:3}, respectively. It is observed that the transverse pressure $ p_\perp $ is less than the radial pressure for $ r $ in the range $ 0 < r < 2.78903. $ Subsequently $ p_\perp $ dominates $ p_r $ in the region $ 2.78903 < r \leq 9.1 $. The radial pressure $ p_r $ vanishes at $ r = 9.1. $ In Figure ~\ref{fig:4}, we have shown the variation of anisotropy throughout the distribution. The variations of sound speed in the radial and transverse directions are shown in Figure ~\ref{fig:5}. From Figure ~\ref{fig:6}, it is evident that the strong energy condition, $ \rho - p_r - 2p_\perp \geq 0 $ is satisfied throughout the distribution. Though we have not assumed any explicit expression for the EOS in our model, we have shown the nature of variation of pressures $ p_r $ and $ p_\perp $ against density in Figure ~\ref{fig:7}. For a relativistic model to be stable in its region of validity, we must have the adiabatic index $ \Gamma > \frac{4}{3}. $ The variation $ \Gamma $ against radius is shown in Figure ~\ref{fig:8}. It is clear from Figure~\ref{fig:8} that $ \Gamma > \frac{4}{3} $ throughout the star. The variation of gravitational red shift, $ z(r) = \sqrt{e^{-\nu(r)}}-1 $ in the radial direction is shown in Figure~\ref{fig:9}. It is easy to note that the red shift is monotonically decreasing function from the centre to boundary. Further, the red shift at the centre $ z_c $ and on the boundary $ z_R $ are both positive and finite. \\

\pagebreak

\pagebreak

\pagebreak

\section{Discussion}
\label{sec:5}

\noindent Spherical distribution of matter on pseudo-spheroidal spacetimes have been studied by a number of researchers in the recent past ~\cite {Tikekar98,Tikekar99,Tikekar05,Thomas05,Thomas07,Paul11,Chattopadhyay10,Chattopadhyay12}. In this paper, we have obtained a new class of solutions to Einstein's field equations for a spherically symmetric anisotropic distribution of matter and have shown that our model can fit to the observational data of a number of well studied pulsars (~\cite {Gangopadhyay13}). On assuming a particular form of radial pressure and on the basis of elementary criteria for physical acceptability of a compact spherically symmetric distribution of matter, we have obtained the bounds for the physical as well as geometric parameters of the model. It is found that our model can accommodate a number of pulsars like 4U 1820-30, PSR J1903+327, 4U 1608-52, Vela X-1, PSR J1614-2230, SMC X-4 and Cen X-3. We also have studied, in detail, a particular pulsar 4U 1820-30, and have shown graphically the profile of different physical quantities throughout the distribution. In short, study of compact stars on the background of pseudo-spheroidal spacetime is highly interesting in the sense that it generates models compatible with observational data and at the same time having a definite 3-space geometry, namely, pseudo-spheroidal geometry which many other spacetimes may not possess.\\

\section*{Acknowledgements}

The authors would like to thank IUCAA, Pune for the facilities and hospitality provided them for carrying out this work. \\


\begin{thebibliography}{99}

\bibitem[\protect \citeauthoryear{Murad}{2013a}]{Murad13a} Murad M. H., {\it Astrophys. Space Sci.} {\bf343} (2013) 187. {\url{doi: 10.1007/s10509-012-1258-4}}.
\bibitem[\protect \citeauthoryear{Murad and Saba}{2013b}]{Murad13b} Murad M. H. and Saba F., {\it Astrophys. Space Sci.} {\bf343} (2013) 587. {\url{doi: 10.1007/s10509-012-1277-1}}.
\bibitem[\protect \citeauthoryear{Murad and Saba}{2013c}]{Murad13c} Murad M. H. and Saba F., {\it Astrophys. Space Sci.} {\bf344} (2013) 69. {\url{doi: 10.1007/s10509-012-1320-2}}.
\bibitem[\protect \citeauthoryear{Murad and Saba}{2014a}]{Murad14a} Murad M. H. and Saba F., {\url{arXiv:1408.5126v2}} (2014).
\bibitem[\protect \citeauthoryear{Maurya \emph{et al.}}{2015}]{Maurya15} Maurya S. K., Gupta Y. K., Ray S. and Chowdhury S. R. {\url{arXiv:1506.02498v1}} (2015).
\bibitem[\protect \citeauthoryear{Sharma and Ratanpal}{2013}]{Sharma13} Sharma R. and Ratanpal B. S., {\it Int. J. Mod. Phys. D} {\bf13} (2013) 1350074.{\url{doi: 10.1142/S0218271813500740}}.	
\bibitem[\protect \citeauthoryear{Vaidya and Tikekar}{1982}]{Vaidya82} Vaidya P. C. and Tikekar R. , {\it J. Astrophys. Astron.} {\bf3}, (1982) 325. {\url{doi: 10.1007/BF02714870}}.
\bibitem[\protect \citeauthoryear{Tikekar}{1990}]{Tikekar90} Tikekar R. S., {\it Journal of Mathematical Physics} {\bf31} (1990) 2454. {\url{doi: 10.1063/1.528851}}.
\bibitem[\protect \citeauthoryear{Finch and Skea}{1989}]{Finch89} Finch M. R. and Skea J. E. F., {\it Class. Quantum Grav.} {\bf6} (1989) 467. {\url{doi: 10.1088/0264-9381/6/4/007}}.
\bibitem[\protect \citeauthoryear{Tikekar and Jotania}{2005}]{Jotania05}Tikekar R. and Jotania K., {\it Int. J. Mod. Phys. D} {\bf14} (2005) 1037. {\url{doi: 10.1142/S021827180500722X}}.
\bibitem[\protect \citeauthoryear{Tikekar and Thomas}{1998}]{Tikekar98} Tikekar R. and Thomas V. O., {\it Pramana J. Phys.} {\bf50} (1998) 95. {\url{doi: 10.1007/BF02847521}}.
\bibitem[\protect \citeauthoryear{Ruderman}{1972}]{Ruderman72} Ruderman R., {\it Astro. Astrophys.} {\bf10} (1972) 427. {\url{doi: 10.1146/annurev.aa.10.090172.002235}}
\bibitem[\protect \citeauthoryear{Canuto}{1974}]{Canuto74} Canuto V., {\it Annu. Rev. Astron. Astrophys.} {\bf12} (1974) 167. {\url{doi: 10.1146/annurev.aa.12.090174.001123}}.
\bibitem[\protect \citeauthoryear{Bowers and Liang}{1974}]{Bowers74} Bowers R. and Liang E., {\it Astrophys. J.} {\bf188} (1974) 657. {\url{doi: 10.1086/152760}}.
\bibitem[\protect \citeauthoryear{Maharaj and Maartens}{1989}]{Maharaj89} Maharaj S. D. and  Maartens R., {\it Gen. Relativ. Grav.} {\bf21} (1989) 899. {\url{doi: 10.1007/BF00769863}}.
\bibitem[\protect \citeauthoryear{Gokhroo and Mehra}{1972}]{Gokhroo94} Gokhroo M. K. and Mehra A. L., {\it Gen. Rel. Grav} {\bf26} (1994) 75. {\url{doi: 10.1007/BF02088210}}.
\bibitem[\protect \citeauthoryear{Tikekar}{1999}]{Tikekar99} Tikekar R.  and Thomas V. O., {\it Pramana J. Phys.} {\bf52} (1999) 237. {\url{doi: 10.1007/BF02828886}}.
\bibitem[\protect \citeauthoryear{Tikekar and Thomas}{2005}]{Tikekar05} Tikekar R. and Thomas V. O., {\it Pramana J. Phys.} {\bf64} (2005) 5. {\url{doi: 10.1007/BF02704525}}.
\bibitem[\protect \citeauthoryear{Thomas \emph{et al.}}{2005}]{Thomas05} Thomas V. O., Ratanpal B. S. and Vinodkumar P. C., {\it Int. J. Mod. Phys. D} {\bf14} (2005) 85. {\url{doi: 10.1007/s12043-012-0268-7}}.
\bibitem[\protect \citeauthoryear{Thomas and Ratanpal}{2007}]{Thomas07} Thomas V. O. and Ratanpal B. S., {\it Int. J. Mod. Phys. D} {\bf16} (2007) 9. {\url{doi: 10.1142/S0218271805005852}}.
\bibitem[\protect \citeauthoryear{Dev and Gleiser}{2002}]{Dev02} Dev K. and Gleiser M., {\it Gen. Rel. Grav.} {\bf34} (2002) 1793. {\url{doi: 10.1023/A:1020707906543}}.
\bibitem[\protect \citeauthoryear{Dev and Gleiser}{2003}]{Dev03} Dev K. and Gleiser M., {\it Gen. Rel. Grav.} {\bf35} (2003) 1435.10. {\url{doi: 1023/A:1024534702166}}.
\bibitem[\protect \citeauthoryear{Dev and Gleiser}{2004}]{Dev04} Dev K. and Gleiser M., {\it Int. J. Mod. Phys. D} {\bf13} (2004) 1389. {\url{doi: 10.1142/S0218271804005584}}.
\bibitem[\protect \citeauthoryear{Sharma and Maharaj}{2007}]{Sharma07} Sharma R. and Maharaj S. D., {\it Mon. Not. R. Astron. Soc.} {\bf375} (2007) 1265. {\url{doi:  10.1111/j.1365-2966.2006.11355.x}}.
\bibitem[\protect \citeauthoryear{Thirukkanesh and Maharaj}{2008}]{Thirukkanesh08} Thirukkanesh S.  and Maharaj S. D., {\it Class. Quantum Grav.} {\bf25} (2008) 235001. {\url{doi: 10.1088/0264-9381/25/23/235001}}.
\bibitem[\protect \citeauthoryear{Komathiraj and Maharaj}{2007}]{Komathiraj07} Komathiraj K. and Maharaj S. D., {\it Intenational Journal of Modern Physics D} {\bf16} (2007) 1803. {\url{doi: 10.1142/S0218271807011103}}.
\bibitem[\protect \citeauthoryear{Sunzu \emph{et al.}}{2014}]{Sunzu14} Sunzu J. M., Maharaj S. D., Ray S., {\it Astrophys. Space Sci.} {\bf352} (2014) 719. {\url{doi: 10.1007/s10509-014-1918-7}}.
\bibitem[\protect \citeauthoryear{Feroze and Siddiqui}{2011}]{Feroze11} Feroze T. and Siddiqui A. A., {\it Gen. Relativ. Grav.} {\bf43} (2011) 1025. {\url{doi: 10.1007/s10714-010-1121-2}}.
\bibitem[\protect \citeauthoryear{Maharaj and Takisa}{2012}]{Maharaj12} Maharaj S. D. and Takisa P. M., {\it Gen. Relativ. Grav.} {\bf44} (2012) 1419. {\url{doi: 10.1007/s10714-012-1347-2}}
\bibitem[\protect \citeauthoryear{Varela \emph{et al.}}{2010}]{Varela10} Varela V., Rahaman F., Ray S., Chakraborty K. and Kalam M., {\it Phys. Rev. D} {\bf82} (2010) 044052. {\url{doi: 10.1103/PhysRevD.82.044052}}.
\bibitem[\protect \citeauthoryear{Paul \emph{et al.}}{2011}]{Paul11} Paul B. C., Chattopadhyay P. K., Karmakar S. and Tikekar R., {\it Mod. Phys. Lett. A} {\bf26} (2011) 575. {\url{doi: 10.1142/S0217732311034943}}.
\bibitem[\protect \citeauthoryear{Thirukkanesh and Ragel}{2012}]{Thirukkanesh12} Thirukkanesh S., Ragel F. S., {\it Pramana J. Phys.} {\bf78} (2012) 687. {\url{doi: 10.1007/s12043-012-0268-7}}.
\bibitem[\protect \citeauthoryear{Maharaj and Takisa}{2013b}]{Maharaj13b} Maharaj S. D. and Takisa P. M., {\it Gen. Relativ. Grav.} {\bf45} (2013b) 1951. {\url{doi: 10.1007/s10714-013-1570-5}}.
\bibitem[\protect \citeauthoryear{Malaver}{2013a}]{Malaver113} Malaver M., {\it American Journal of Astronomy and Astrophysics} {\bf1} (2013) 41. {\url{doi:  10.11648/j.ajaa.20130104.11}}.
\bibitem[\protect \citeauthoryear{Malaver}{2013b}]{Malaver213} Malaver M., {\it World Applied Programming } {\bf3} (2013) 309.
\bibitem[\protect \citeauthoryear{Malaver}{2014}]{Malaver14} Malaver M., {\it Frontiers of Mathematics and its Applications } {\bf1} (2014) 9. {\url{doi: 10.12966/fmia.03.02.2014}}.
\bibitem[\protect \citeauthoryear{Thirukkanesh and Ragel}{2014}]{Thirukkanesh14} Thirukkanesh S., Ragel F. S., {\it Pramana J. Phys.} {\bf83} (2014) 83. {\url{doi: 10.1007/s12043-014-0766-x}}.
\bibitem[\protect \citeauthoryear{Pandya \emph{et al.}}{2015}]{Pandya14} Pandya D. M., Thomas V. O., Sharma R., {\it Astrophys. Space Sci.} {\bf356} (2015) 285. {\url{doi: 10.1007/s10509-014-2207-1}}.
\bibitem[\protect \citeauthoryear{Gangopadhyay \emph{et al.}}{2013}]{Gangopadhyay13} Gangopadhyay T., Ray S., Li X-D., Dey J. and Dey M., {\it Mon. Not. R. Astron. Soc.} {\bf431} (2013) 3216. {\url{doi: 10.1093/mnras/stt401}}.
\bibitem[\protect \citeauthoryear{Chattopadhyay and Paul}{2010}]{Chattopadhyay10} Chattopadhyay P. C. and Paul B. C., {\it Pramana- j. of phys.} {\bf74} (2010) 513. {\url{doi:  10.1007/s12043-010-0046-3}}.
\bibitem[\protect \citeauthoryear{Chattopadhyay \emph{et al.}}{2012}]{Chattopadhyay12} Chattopadhyay P. C., Deb R. and Paul B. C., {\it International Journal of Modern Physics D} {\bf21} (2012) 1250071. {\url{doi:10.1142/S021827181250071X}}.

\end{thebibliography}
\end{document}